\documentclass[letterpaper,11pt]{article}
\usepackage[dvips]{graphicx}
\title{Accuracy of an Atomic Microwave Power Standard} 
\author{David C. Paulusse, Nelson L. Rowell, and Alain Michaud\footnote{The authors are with the Institute for National Measurement Standards, National Research Council, Ottawa, ON., Canada, K1A 0R6. email: Alain.Michaud@nrc-cnrc.gc.ca. This paper was submitted for publication in IEEE Trans. Instrum. Meas. }}

\date{July 2, 2004, revised October 10, 2004}

\begin{document}

\maketitle

\paragraph{Abstract--}
We have built an atomic microwave power standard based on the electromagnetic interaction with laser-cooled atoms. The atoms traversed a waveguide transmission line, and under the effect of the radiation, the internal state populations underwent a Rabi flopping oscillation. Measurement of the oscillation frequency allowed the determination of the incident microwave power. As many of 60 oscillations were observed over a dynamic range of 20 dB and the standard deviation of the measurements was about 0.02\%.  The measured frequency was compared to a calculated one and an agreement of   1.3\% with an uncertainty of 5\% (rectangular) was found.



\begin{centering}
\section*{\sc Introduction}
\end{centering}

Although the standard calorimetric technique is a proved and practical way to measure microwave power it is still worthwhile to develop alternate techniques, e.g. ones based on the probing of atomic transitions. Similar and different approaches have been investigated in the past and are described in  \cite{fantom}. The advent of laser cooling brings more interesting potentialities for future developments. Two experiments have already demonstrated that the measurement of the Rabi flopping frequency of laser cooled atoms can be related to the microwave power. Our previous prototype had a resolution of 0.3\% over a dynamic range of 20dB \cite{cpem2002conf}, \cite{cpem2002paper}. As the radiation was applied to the atoms through a glass cell, it was not possible to evaluate the field distribution nor to accurately measure the power. The National Institute of Standards and Technology experiment consisted of a launched cloud of atoms through a resonant cavity \cite{crowley}. Although the high quality factor of the cavity made it difficult to accurately evaluate the incident power, they reported agreement between the measured power using the Cs atoms and a power meter within about 4\%.

In this paper, the atoms crossed an actual waveguide transmission line that allowed a more accurate description of the electromagnetic field and its relation with the transmitted power. We give a description of the apparatus, show a typical measurement, and discuss its accuracy.
\\

\begin{centering}
\section*{\sc II. Principle of Operation}
\end{centering}

The model that describes the evolution of the atomic states when they are submitted to electromagnetic field was developed for the study of the atomic clocks. However if the frequency of the field is kept in resonance with the transition, then the atomic populations will undergo a Rabi flopping oscillation whose frequency is proportional to the amplitude of the electromagnetic field (see \cite{vanieraudoin} chapter 5). In order to achieve an accurate measurement, one should evaluate the field amplitude on the path of the atoms and should also should know the ratio between the power measured by the atoms in vacuum to the  incident power in air.

This experiment used laser cooled-Rubidium atoms, which were dropped through a waveguide transmission line operating in the fundamental mode (See Fig. 1). The advantages of this scheme are that the microwave field distribution is very well known and the transmitted power can be calculated using an analytical expression. The waveguide section was kept under vacuum using two reflectionless windows and the transmitted power could be measured simultaneously via the cold atoms and a monitor power sensor.  The device was operated in CW mode and is fully compatible with common waveguide standards.
\\

\begin{figure}
\includegraphics[scale=1.2,bb= 180 250 450 500]{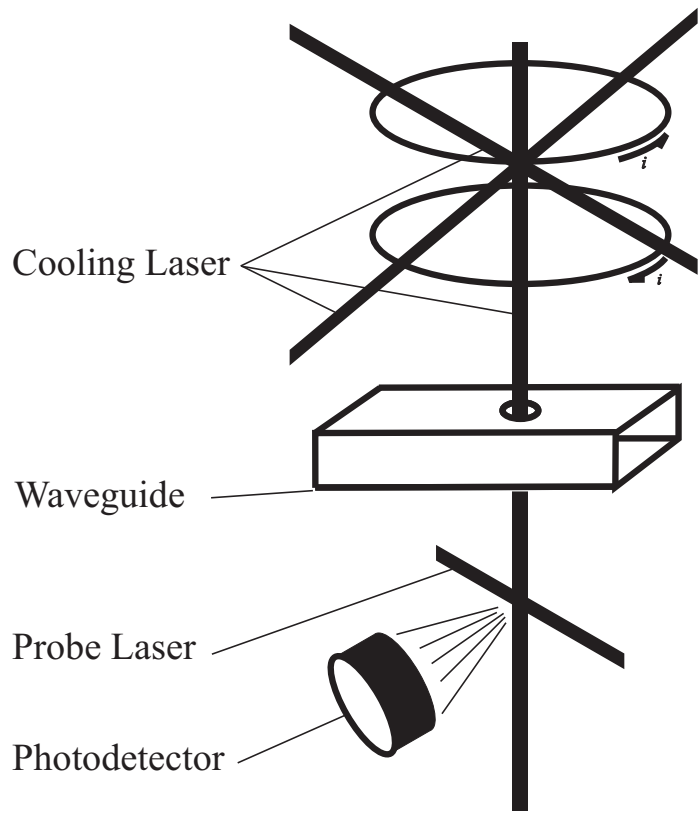}
\caption{Schematic of the experimental system. The atoms are trapped at the intersection of the laser beams, then they cross the waveguide and are detected below.}
\end{figure}

\subsubsection*{A. Lasers and Optics}

The laser system was based on single frequency, stabilized laser diodes. The cooling laser delivered about 35 mW and was kept in resonance with $5s_{1/2}$ F= 2 to $5p_{3/2}$ F'=3 transition of Rb$^{87}$. The probe and repumping lasers, were tuned to the same transition as the cooling laser and to the  F=1 to the F'=(1 or 2) transition respectively.

First, approximately 10$^{8}$ atoms were captured in a standard magneto-optical trap \cite{metcalf}. Next, the atoms were optically pumped simply by turning off the repumping light earlier than the cooling laser at the end of the cooling sequence. At this point all the atoms were in the F=1 state.

Before they entered the waveguide, a static magnetic field of 160 milligauss parallel to the magnetic part of the  RF field was ramped up in about 30 ms. The atoms crossed the guide in 14 ms and then they were probed using a laser beam. The fluorescence signal from the photodetector measured the population of the F=2 level, so any signal results from the m$_{F}$=0 to m$_{F'}$=0 transition only.
\\

\subsubsection*{B. Microwave System}

The atomic power standard is compared to a power transfer standard which was operated in the calibrated source mode. The power transfer standard was a 3 dB waveguide directional coupler and a NBS-IV type power meter. Its output was attached to one end of the transmission line vacuum chamber. The other end of the line was attached to a reflection-less termination.

In this experiment the microwave power was kept constant (CW  mode) and the field intensity time profile was set by the time of flight of the atoms through the waveguide. We repeated the experiment varying the power level each time and we evaluated the number of oscillations. Knowing the analytical expression for the magnetic field \cite{seshadri}, we could deduce the calibration factor $K2$, the ratio of incident power to the monitored power on the side arm of the transfer standard \cite{fantom}.

The reference frequency input of the synthesizer was obtained from an atomic clock and the output frequency was set to the atomic transition frequency. Since the Rabi frequency is also a function of the detuning from the resonance, it was necessary to set the signal frequency with care. The microwave signal was generated by the synthesizer was also leveled using a 10 dB coupler and thermistor mount-power meter (coaxial).
\\

\subsubsection*{C. Evacuated Transmission Line}

The vacuum chamber consisted of an evacuated R-70 type waveguide transmission line in which the fundamental mode propagates to the interaction region without reflection or leakage. We prevented any reflection by using tuned windows on each end of the line (Fig. 2). Each window consisted of two 1 mm thick quartz plates spaced a quarter wavelength apart. The reflection from the first plate canceled exactly the reflection of the second one and the insertion loss of the window was therefore small at the frequency of operation. Furthermore, as an additional precaution, against any possible standing waves between the windows, the distance between them was chosen so that the frequency of operation is centered between the two closest modes.  The overall length of the line is about 50 cm, and it is symmetrical around the access holes.

\begin{figure}[t]
\includegraphics[width=4in,bb= 150 350 400 400]{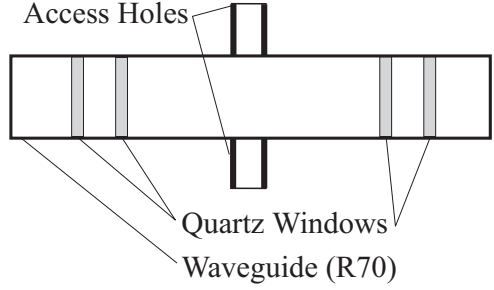}
\caption{Diagram of the evacuated transmission line. The device is reversible.}
\end{figure}

Fig 3 shows the overall insertion loss of the line. The vertical line on the graph marks the operating frequency. The baseline of 1.05\% corresponds to the normal attenuation of the copper waveguide. The oscillations on both sides of the operating frequency are interference between the \"input\" and \"output\" windows. They disappear in the center as the windows are tuned not to have any reflection at that frequency. The access holes are located so that the atoms will cross the guide in the center of the broad walls. This position was chosen to obtain linear magnetic field polarization and constant field intensity in the path of the atoms. The holes are 5.7~mm in diameter and 12~mm long with negligible measured leakage. A more accurate evaluation of the interaction profile would include a numerical simulation of this area but this was not done in the present study. Finally, using laser beam geometry, which would not use a vertical laser beam could further reduce the diameter of the holes.

\begin{figure}
\includegraphics[width=3in,bb= 100 50 550 200, angle=-90]{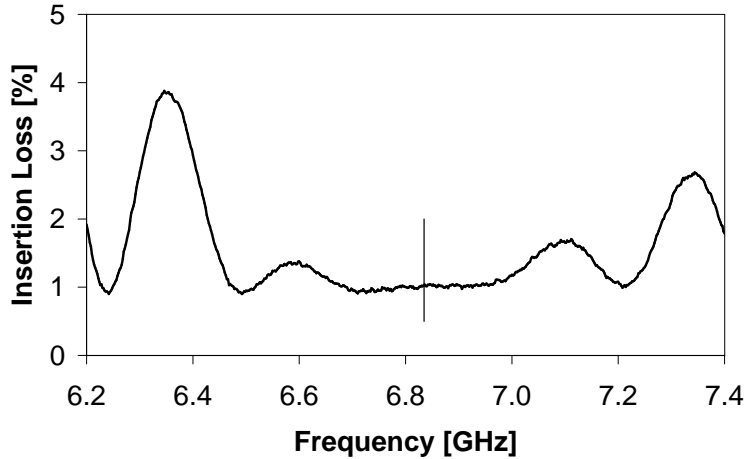}
\caption{Frequency response of the transmission line. The vertical line marks the operating frequency.}
\end{figure}

At approximately 14 cm from the center of the line, there was also a grid on the narrow wall of the waveguide (not shown in the figure). The purpose of that grid was to allow the high vacuum pumping for the waveguide itself.

The insertion loss measurement were done using a vector network analyzer. We also measured a line of the same guide having the same length. The measured insertion loss of this line was found to be 0.8\% ($\pm$ 0.1\%). The 0.25\% difference between this and the chamber is the sum of all the contributions: the tuned windows, the access holes where the atoms crosses the chamber and the grid for the vacuum pumping. 
\\

\begin{centering}
\section*{\sc III. Experimental results and Discussion}
\end{centering}

\subsubsection*{A. Measurement Results}

The normalized amplitude of the photodetector signal is plotted in Fig. 4, as a function of the square root of the incident power.  The solid line is a fit of a cosine function.  The number of samples was chosen to unambiguously resolve the number of fringes in a short amount of time. The measurements were done as follow: Three \"dry\"  launches were done as a check of the trap quality. Then one launch was done with the optical pumping applied. Typically the signal fell to 10 percent of the previous amplitude. Finally the microwave power was turned on, the amplitude stabilization loop left to settle and three launches were done with the optical pumping and microwave present in the waveguide. The plotted value is normalized with respect to the difference between  the no-microwave and no-pumping amplitudes. Then the DC power applied on the monitor thermistor mount was measured both with and without RF applied. The normalization of the vertical axis was done in order to reduce the noise and the effect of long term drifts such as trap size for example.  

\begin{figure}[t]
\includegraphics[bb=100 50 599 780,angle=-90,width=5.5in]{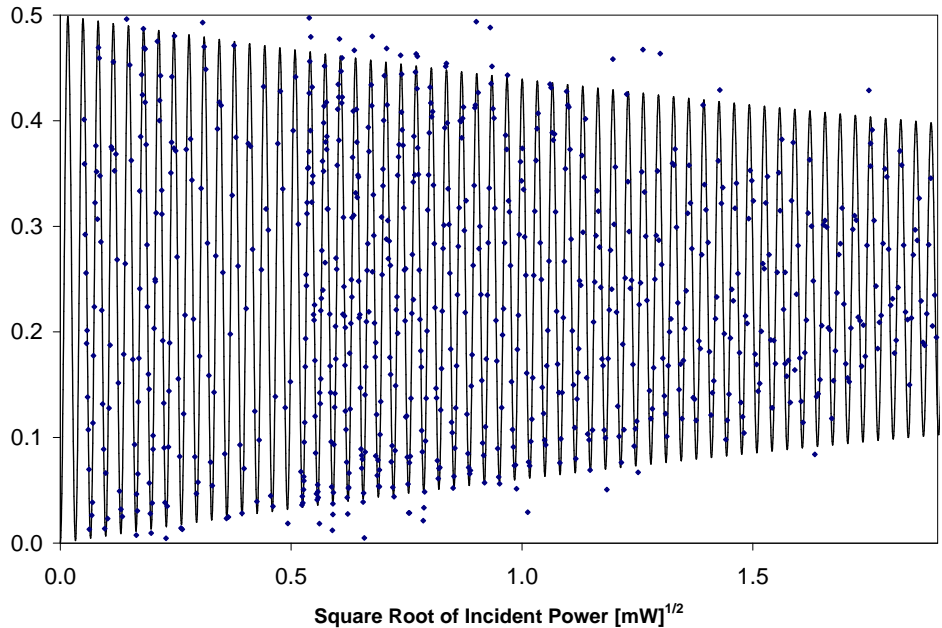}
\caption{Normalized amplitude of the photodetector signal as a function of the square root of transmitted power.}
\end{figure}

Fig 4 shows as many as 60 Rabi oscillations over a power range up to 3.9 mW. This range includes the usual measurement conditions commonly employed, i.e. 1 mW. Higher ranges are possible simply by launching from a higher distance.  The loss of contrast at the high range of the curve was from three sources: 1) The atoms travel at different velocity due to the fact that the temperature is not zero. The phase difference between the slow and the fast atoms increases as the average frequency increases thus resulting in a loss of contrast. 2) The field intensity was maximum but not constant along the large wall direction at the center of the guide. 3) The sinusoidal behavior is given under the rotating field approximation. Since the only important parameter in this figure is the frequency of the oscillations, no attempt was done to characterize the amplitude of the oscillations or the contrast decay. This is usually achieved using a small scale (classical) ballistic monte-carlo analysis. All the atom's contribution are summed, assuming an initial position and velocity distribution (see the references in \cite{metcalf} section 7.4).  This was not done in the present study. Nevertheless, measurements over the whole range are not necessary if an estimate of the power is known. Then a certain number of selected fringes could be used only, possibly the last ones. 
\\

\subsubsection*{B. Linearity and Noise }

In order to evaluate the linearity of the system, we compared a short section of the data plotted in Fig. 4 to the full sequence.  Each relative deviation $R_{i}$ plotted in Fig. 5 is obtained by fitting three cycles of data (about 30 points) to the function: 

\begin{equation}
	S(P) \simeq  \cos ( \Omega \sqrt{P} (1 + R_{i}))
\end{equation}
\\
\begin{figure}[t]
\includegraphics[bb= 50 50 550 650, angle=-90,width=4in]{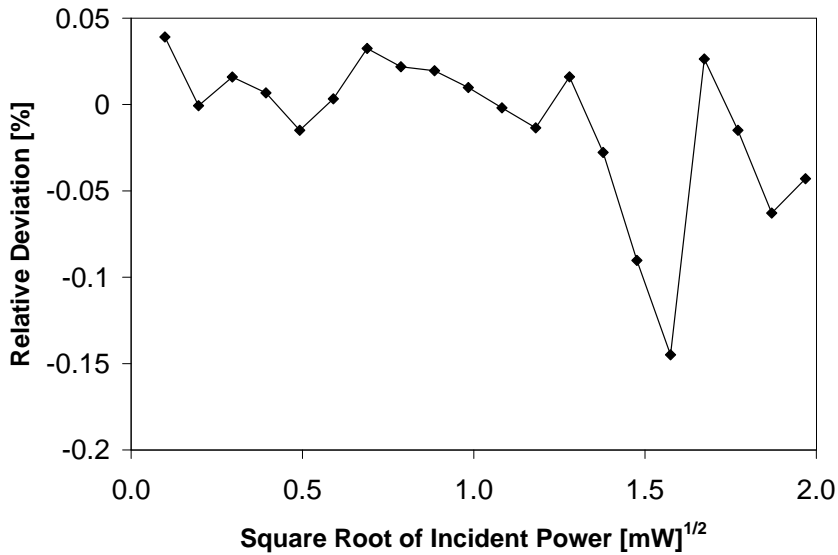}
\caption{Relative deviation of the Rabi frequency versus a linear dependance.}
\end{figure}

We observed a dip at around 2 mW where the deviations were as large as 0.14\%.  It is not clear at this stage whether this was due to a non-linear effect or it was merely noise. If we restricted the measurement range to less than 1 mW then the deviation was typically around    0.02\% or less, corresponding to an averaging time of 10 to 30 minutes per point (3 cycles of data). 
\\
\subsubsection*{C. Accuracy} 

We calibrated the transfer standard by comparison with an existing PC-7 mm standard. A waveguide to coaxial adapter was used to attach a waveguide thermistor mount to the calibrated source. The S-parameters of the adapter were measured on a vector network analyzer and used in the expression of the correction factor. This resulted in accuracy for the power measurements of 2\% with a rectangular distribution.

In order to evaluate the interaction time, we measured the distance from the trap to the waveguide and to the probe beam and compared them to the observed values. We also compared our estimation to the time of flight signal.  Finally, we estimated the relative uncertainty on the interaction time to be 4\% (rectangular). This represents the most important source of uncertainty of our experiment. 

Now we compare the measured Rabi frequency  from the figure 4 to the one deduced from the measurement of the incident power we find:
\begin{equation}
\frac{\Omega_{M}}{\Omega_{C}} = 0.987 \pm 0.05
\end{equation}

When measuring power, both techniques were in agreement by 2.6\%. The total uncertainty in power was less than 10\% (rectangular distribution) and was mainly due to the transit time (4\%) and the reference power sensor (2\%). This suggests that more tests should be made in order to further reduce the uncertainties. However, although we can not say at this time that the accuracy of the present method has reached that of an actual standard, we have shown that the system allows one to perform measurements with a resolution of 0.02\% over a range of 20 dB with no obvious fundamental limitation, to the accuracy.\\

\begin{centering}
\section*{\sc IV. Conclusion and Outlook}
\end{centering}

We have measured the Rabi frequency of the hyperfine interaction of a laser cooled sample of Rb$^{87}$ atoms as they fall through a microwave transmission line. It was possible to observe as many as 60 flopping oscillations up to a power of 3.9 mW, corresponding to a dynamic range of 26 dB. The frequency of the oscillations was measured with a resolution of 0.02\%.  It was compared to the calculated frequency and an agreement of 1.3\% was found between both, with an accuracy of 5\% (rectangular distribution). This corresponds to an agreement of 2.6\% when considering power measurement. The accuracy of our system could significantly be improved by reducing the uncertainty of the transit time through the waveguide. As there are many ways by which this objective can be reached, it therefore does not constitute a major limitation to realizing this type of standard.
\\


\begin{thebibliography}{9}

\bibitem{fantom} \textsc{A.~Fantom}. \emph{Radio Frequency and Microwave Power Measurement},	 \'Ed. Peter Peregrinius Ltd., London, 278 pp.,  1990, IEE press, Available: http://www.iee.org

\bibitem{cpem2002conf} \textsc {D.C.~Paulusse, N.L.~Rowell and A.~Michaud}, Realization of an Atomic Power Standard, 2002 Conference on Precision Electromagnetic Measurements, Ottawa, June 16-21, 2002, pp. 194-5.

\bibitem{cpem2002paper} \textsc{D.C.~Paulusse, N.L.~Rowell, A.~Michaud}, "Microwave Power Standard using cold Atoms", Eighteenth International Conference on Atomic Physics (ICAP 2002) Poster Presentation Abstracts.	Cambridge, July 28-August 2, 2002, p. 329[Online]. Available: http://michaud.jdhosts.net/a2002/2423055.pdf, also http://arxiv.org/physics/0503120 

\bibitem{crowley} \textsc{E.A.~Donley, T.P.~Crowley, T.P.~Heavner, B.F.~Riddle},"Quantum-Based Microwave Power Measurement Performed with a Miniature Atomic Fountain", Proceedings: 2003 IEEE International frequency Control Symposium.	Tampa Bay, May 4-8, 2003, pp. 135–137.

\bibitem{vanieraudoin} \textsc{J.~Vanier, C.~Audoin}, \emph{The Quantum Physics of Atomic Frequency Standards}. Adam Hilger, Bristol, 1988, Vol. 1,2. 1567 pp.

\bibitem{metcalf} \textsc{H.J.~Metcalf, P.~van~der~Straten}, \emph{Laser Cooling and Trapping}	New York, Springer-Verlag, 1999, pp. 156–162.

\bibitem{seshadri} \textsc{S.R.~Seshadri}, \emph{Fundamentals of Transmission Lines and Electromagnetic Fields}	Addison-Wesley, Massachusetts, 1971, 598 pp.

\end{thebibliography}
\end{document}